# Suppression of superconductivity in nanowires by bulk superconductors

Mingliang Tian, Nitesh Kumar, Shengyong Xu, Jinguo Wang, James S. Kurtz, and M. H.W. Chan
*The Center for Nanoscale Science and Department of Physics,*
*The Pennsylvania State University, University Park, Pennsylvania 16802-6300*

Transport measurements were made on a system consisting of a zinc nanowire array sandwiched between two bulk superconducting electrodes (Sn or In). It was found that the superconductivity of Zn nanowires of 40 nm diameter is suppressed either completely or partially by the superconducting electrodes. When the electrodes are driven into their normal state by a magnetic field, the nanowires switch back to their superconducting state. This phenomenon is significantly weakened when one of the two superconducting electrodes is replaced by a normal metal. The phenomenon is not seen in wires with diameters equal to and thicker than 70 nm.

When a normal metal (N) is placed in contact with a superconductor (S), it acquires superconducting property over a characteristic length, typically of a few micrometers. This is known as the proximity effect and has been extensively studied in planar SN junctions and in metallic strips lithographically patterned between two superconductors [1-3]. Related behavior is found in a constriction system consisting of a narrow strip of superconductor with a length L, bridging two macroscopic superconducting electrodes of the same material. If L is much shorter than the coherence length, $\xi$, the critical current was found to be enhanced as compared to that in an isolated long wire (i.e., L >> $\xi$) of the same cross section [1, 4].

In this letter, we report a novel and unexpected "anti-proximity" effect in a system BS/ZNW/BS comprising of superconducting zinc nanowires (ZNW's) sandwiched between two bulk superconductors (BS's) of different materials (Sn, and In). We found evidences that the superconductivity of ZNW's of 40 nm in diameter is suppressed either completely or partially when the BS's are in the superconducting state. When the BS's are driven into their normal state by a magnetic field, the nanowires switch back to their superconducting state.

Bulk zinc (Zn) is a conventional type-I superconductor with a transition temperature of 0.85 K (at H = 0 Oe) and a critical magnetic field of 50 Oe (at T = 0 K). We chose Zn because its superconducting coherence length at T = 0 K, $\xi(0)$, is estimated to be ~ 1.5-2.2 $\mu$m [5], thus the one-dimensional (1d) characters in a Zn nanowire may be more apparent than in other materials with smaller coherence length. The nanowires are electrochemically deposited into the 1d channels of commercially available porous polycarbonate (PC) or porous alumina membranes [6]. The electrolyte was prepared by dissolving 4.7 g $ZnCl_2$ into 200 ml distilled water, and then mixed with 40 ml saturated KCl and 0.5g gelatin. Prior to the electrodeposition, a 200 nm Au film was evaporated onto one side of the membrane that served as the cathode. A pure bulk Zn wire was used as the anode. The deposition of ZNW's was carried at room temperature under a voltage that ranges from -0.1 to -0.4 V. Structural characterization of the ZNW's released from a membrane was made by field-emission transmission electron microscopy (FETEM). The TEM images and the electron diffraction pattern shown in Fig. 1(a)-(b) indicate the majority of the nanowires has polycrystalline structure with elongated crystalline segments of several hundred nanometers in length.

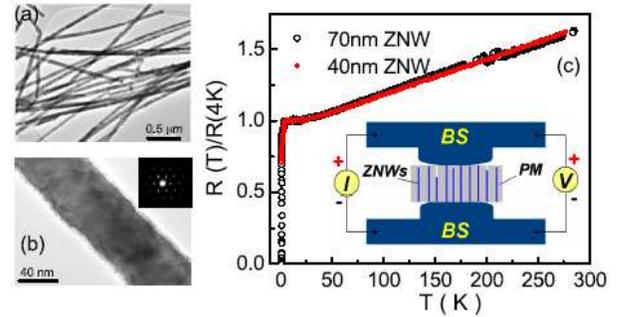

FIG. 1: TEM images of (a) freestanding 40 nm Zn wires and (b) a segment showing crystalline structure with [0001] orientation of a wire. (c) Normalized R(T)/R(4K) versus T of 70 and 40 nm ZNW between bulk In electrodes from 0.47 K to 300 K, and the schematic of transport measurement on the BS/ZNW/BS sample.

Electrical transport measurements were carried out with a Physical Properties Measurement System (Quantum Design), which is equipped with a He-3 cryostat and a superconducting magnet. The experimental arrangement is shown schematically in the inset of Fig. 1(c). High-purity (99.9999%) Sn or In wires of 0.5 mm diameter and typically 2 cm in length were mechanically squeezed onto the two sides of a membrane making electrical contact to the nanowires embedded in the membrane [7]. In this configuration, the Sn or In wires are the BS's of the BS/ZNW/BS structure. The BS on each side of the sample bifurcates into two leads, allowing 4 lead measurements on the BS/ZNW/BS system. The measured resistance (R) has contributions from the BS's, ZNW's and two BS/ZNW interfaces. Fig. 1(c) shows the normalized resistance R(T)/R(4K) of two In/ZNW/In samples with the ZNW's of 70 and 40 nm in diameter, measured under H = 0 Oe. These two R-T curves appear to collapse onto each other, showing metallic behavior from room temperature down to the transition temperature (3.4 K) of In. The metallic behavior above 3.4 K shows that our procedure of squeezing the bulk indium wires to membrane is successful in making ohmic contact with some of the nanowires. With this technique, 40 samples of different configuration (i.e., different diameters and length of ZNW's and bulk electrodes of different material) were measured. Reproducible results are always found on the samples of similar configuration. In this letter, we show data of 7 representative samples Z1, Z2, Z3, Z4, Z5, Z6 and Z7. The numbers of ZNW's making contact to the electrodes in these samples are estimated to be 11, 4, 6, 1, 55, 13, and 5 respectively using a resistivity of $\rho$ ~ 18.5 $\mu\Omega$·cm at 4 K [8]. Based on the uncertainties of the length (~ 5 %)and diameter (~ 10 %) of the wires, the estimated numbers are likely to be correct to within 25%.

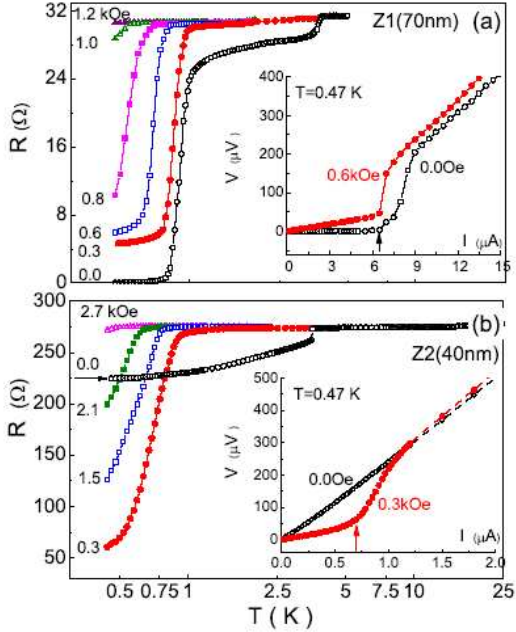

FIG. 2: R-T curves of samples Z1(70 nm) (a) and Z2(40 nm)(b) with the Sn/ZNW/Sn configuration under different magnetic fields aligned perpendicular to the wire axis, measured with an excitation current of I = 0.5 $\mu$A. The insets show the V-I characteristics at H = 0.0 and 0.3 kOe, respectively.

Fig. 2(a) shows the R-T curves of a Sn/ZNW/Sn system with 70 nm diameter ZNW of 6 $\mu$m in length (sample Z1), measured at different magnetic fields. At H = 0 Oe, the resistance of the system shows a small drop near 3.7 K, and then a much larger drop near 1.0 K before decaying to zero within our experimental uncertainty below 0.7 K. The resistance drops at 1.0 K and 3.7 K have their origins in the superconducting transitions of ZNW's and bulk Sn, respectively. In addition to the bulk Sn leads, the observed drop at 3.7 K includes contributions due to the two Sn/ZNW interface regions turning superconducting through the standard proximity effect. The smooth decrease in the resistance from 3.7 K to 1.0 K can be ascribed to the proximity effect of the bulk Sn electrodes extending from the two interface regions into the ZNW's. The fact that zero resistance was found below 0.7 K shows that the entire Sn/ZNW/Sn system is superconducting at low temperature. When bulk Sn is driven to the normal metallic state at H = 0.3 kOe, the measured resistance of the system shows only one large drop at 0.9 K due to the onset of superconductivity of the ZNW's. Although 0.3 kOe is six times the critical field of bulk Zn, the superconducting transition of the ZNW's at 0.9 K is still seen because the critical field of the ZNW's is enhanced by their reduced diameters [9]. The R-T curves at higher fields indicate the critical field of the 70nm diameter ZNW's has been pushed to a value higher than 1.2 kOe below 0.47 K. The finite resistance in the low temperature limit at 0.3 kOe can be due to the resistance of the normal Sn leads and the Sn/ZNW interfaces. The proximity effect described above is not seen at magnetic fields exceeding 0.3 kOe when the Sn electrodes are in the normal state. The inset of Fig. 2(a) shows the V-I measurements at 0.47 K at H = 0 and 0.6 kOe. Regardless of whether the bulk Sn is superconducting or normal, the V-I curves of the system always show a well-defined critical current, $I_c$, near 6.0 $\mu$A, below which the ZNW's are in superconducting state. The linear dependence of V on I seen at I << $I_c$ at H = 0.6 kOe reflects the ohmic nature of the BS/ZNW interfaces and the normal bulk Sn leads.

Fig. 2(b) shows the R-T curves of another Sn/ZNW/Sn sample (Z2) with ZNW's of 40 nm in diameter and 6 $\mu$m in length, measured at different H. The R-T curve at H = 0 Oe shows the expected drop at 3.7 K, the $T_c$ of Sn. However, in strong contrast to the data of Z1, there is no sign of a drop or even a change in the slope near 1.0 K, $T_c$ of the ZNW's. The large finite resistance found at 0.47 K confirms that the ZNW's reside in the "normal" state. When the bulk Sn electrodes are driven to the normal state by a field of 0.3 kOe, a prominent resistance drop is seen at 1.0 K, indicating the superconductivity of the ZNW's is recovered. The finite low temperature resistance at 0.47 K, as in the case of sample Z1, can be attributed to the resistance of the normal Sn leads and the Sn/ZNW interfaces. The results shown in R-T curves are supported by V vs. I scans measured at 0.47 K, as shown in the inset of Fig. 2(b). At H=0 Oe, when the bulk Sn is in superconducting state, the V-I curve of the system shows ohmic linear behavior at all excitation currents. Under a field of 0.3 kOe, when the bulk Sn is in normal state, the V-I curve shows the expected abrupt increase in V near a critical current $I_c$ ~ 0.6 $\mu$A, indicating a transition in ZNW from superconducting to normal state. The linear ohmic section at current below $I_c$, as in Z1, is due to the normal Sn electrodes and Sn/ZNW interfaces. Similar behavior is also observed in single-crystalline 40

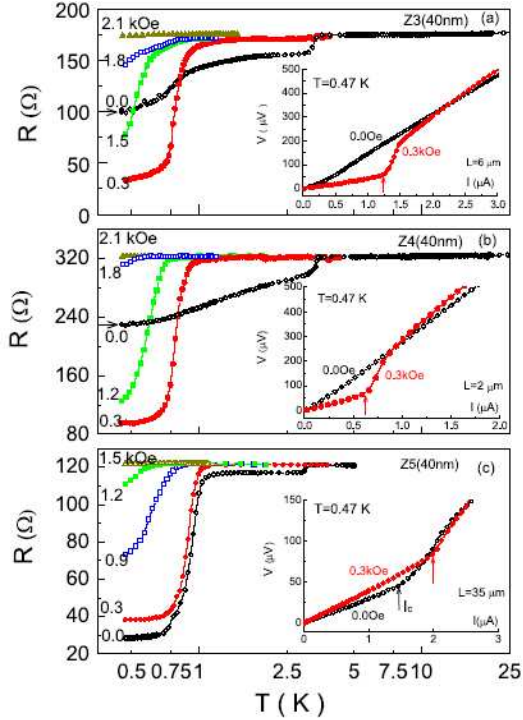

FIG. 3: R-T curves of In/ZNW/In samples with 40 nm ZNW but different lengths (L) under different H aligned perpendicular to the axis of wires measured at I = 10 nA: (a) Z3, L = 6 $\mu$m; (b) Z4, L = 2 $\mu$m; (c) Z5, L = 35 $\mu$m. The insets show the V-I characteristics of the specific sample measured at 0.0 and 0.3 kOe, respectively. The arrows indicate the point of critical current.

nm Zn nanowires with the same configuration.

In order to ensure that the phenomenon shown in Fig. 2(b) is not unique to the Sn/ZNW/Sn system and to understand if the effect depends on the length of the ZNW's, measurements were carried out on three In/ZNW/In samples with ZNW's of 40nm diameter but different lengths. The length of the ZNW in samples Z3 and Z4 is L = 6 $\mu$m and 2 $\mu$m, respectively. The ZNW's in Z5, in contrast to the other 6 samples reported in this paper are made in a porous alumina membrane instead of PC membrane. The length of the wires is 35 $\mu$m. Their R-T curves are respectively shown in Fig. 3(a), (b) and (c). Sample Z3 at H = 0 Oe shows a broad transition near 1.0 K with a limited resistance drop that ends with a large finite resistance of 100 $\Omega$ at 0.47 K. When the bulk indium electrodes are driven normal by a field of 0.3 kOe, a much larger and sharper resistance drop near 1.0 K is observed. This indicates that the superconducting indium electrodes have weakened the superconductivity of ZNW's in Z3. Very similar but much stronger suppression effect (Fig. 3(b)) is seen in sample Z4 where the ZNW is a single wire of 2 $\mu$m. The measured resistance at H = 0 Oe shows a smooth decrease from 3.4 K to 0.47 K with no resistance drop near 1.0 K. Applying magnetic field of 0.3 kOe results in a sharp resistance drop near 1.0 K. The results in Z4 closely resemble that observed in sample Z2, with nanowires of 6 $\mu$m in length but with Sn as the BS electrodes. The V-I scans in the inset of Fig. 3(a) and 3(b) show behavior similar to that of Fig. 2(b) observed with Sn electrodes.

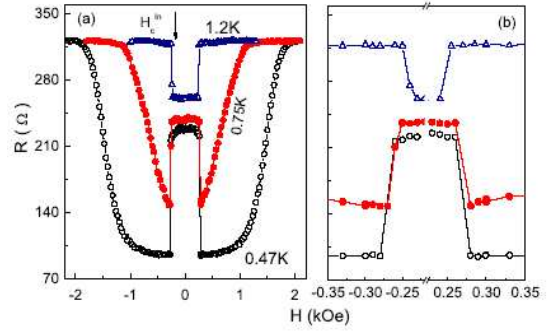

FIG. 4: (a) R-H curves of sample Z4 measured at 0.47 K, 0.75 K and 1.2 K. (b) The magnified part of the R-H curves near the critical field, $\pm H^{in}_c$, of bulk indium.

In contrast to the findings in samples Z3 and Z4, there is no obvious sign of the "anti-proximity" effect in the R-T curves of Z5 (Fig. 3(c)). These curves show a superconducting transition near 1.0 K irrespective of whether the bulk indium electrodes are in the superconducting or normal state. The V-I scans in the inset at 0.47 K measured under two different fields H = 0 Oe and 0.3 kOe, however, do show evidence of the anti-proximity effect. The critical current with the indium electrodes in the normal state (H = 0.3 kOe) is found to be close to 2 $\mu$A, which is higher than the value 1.5 $\mu$A found at 0.0 Oe with indium in superconducting state. The results shown for sample Z3, Z4 and Z5 suggest that the coherence length of Zn, $\xi \sim 2$ $\mu$m, is an important length scale for the observed phenomenon. When the length of the ZNW's, L, is much larger than $\xi$, the effect is found considerably weakened.

Fig. 4(a) shows R-H scans made at 10 nA of sample Z4 (shown in Fig. 3(b)) at different temperatures. The magnified part of the R-H curves near the critical field of indium, $\pm H^{in}_c$, is shown in Fig. 4(b). At 1.2 K, the ZNW's are in the normal state and the discontinuous rise in resistance with increasing field pinpoints the temperature dependent critical field of bulk indium at 0.245 $\pm$ 0.008 kOe. Scans at 0.47 and 0.75 K, below the $T_c$ of bulk Zn, show sharp drops in resistance when the magnetic field is increased to 0.27 $\pm$ 0.01 kOe, the critical field of In at these temperatures. These scans allow us to conclude that the switching of the ZNW's from superconducting to non-superconducting state and vice versa bears a negative correlation with the state, namely superconducting or normal, of the bulk electrodes. The finding of switching of superconductivity in the ZNW's exactly at the critical field of the bulk electrode negates the possibility that the effect is an artifact of electrical noises. Nevertheless, we have carried out control experiments on 40 nm ZNW's of 6 $\mu$m in length sandwiched between two bulk Sn electrodes, with all electrical leads equipped with low pass $\pi$-filters at room temperature and also at low temperatures close to sample.

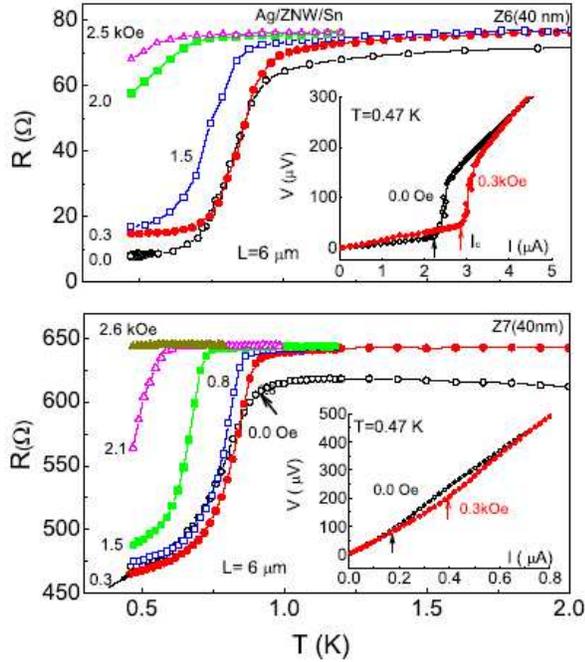

FIG. 5: (a) R-T curves of Z7 (40 nm) with Ag/ZNW/Sn configuration. (b) R-T curves of sample Z6 (40 nm) with Sn as BS and ZNW of 6 $\mu$m in length. The insets show their V-I characteristics at H = 0.0 and 0.3 kOe, respectively.

The anti-proximity effect is reproduced in these measurements with filters.

Transport property of a model system consisting of a mesoscopic superconducting grain coupled by Josephson junctions to two macroscopic superconductors were considered theoretically by Refael and collaborators [10]. They found such a system can go from a phase that is fully superconducting to one that is fully "normal" depending on the value of the total shunting resistance of the two junctions. This model resembles to some degree with our system. There is one important difference. In our case the shunting, or the interface resistance between the nanowires and bulk electrodes as noted above is orders of magnitude smaller than the quantum resistance, $R_q$ = 6.5 k$\Omega$, required in the theoretical model.

In order to understand the effect of BS/ZNW interfaces on the observed phenomenon, control experiments were carried out with two different samples, Z6 and Z7. Sample Z6 has a configuration of Ag/ZNW(40nm)/Sn, with one of the two BS replaced by a normal metal Ag. Sample Z7 has a standard configuration of Sn/ZNW(40nm)/Sn, which is similar to that of Z2 but its interface resistance (~ 460 $\Omega$ at 0.47 K) is 9 times higher than that of Z2. R-T curves for sample Z6 in Fig. 5(a) shows no obvious evidence of anti-proximity effect. The only clear signature of the effect is found in the V-I curves shown in inset of Fig. 5(a). The critical current at H = 0.3 kOe, with the Sn electrode in the normal state is higher than that at zero field. Comparing results of sample Z2 and Z6 shows that the anti-proximity effect is present even when one of the two bulk electrodes is a superconductor. However, it is much stronger when both electrodes are superconducting. It is interesting to note that the result of sample Z6 is similar to the 35 $\mu$m long wire case (sample Z5) with both BS shown in Fig. 3(c). In sample Z7 with much higher interface resistance, as shown in Fig. 5(b), the only signature of the anti-proximity effect in the R-T curves is the slightly lower R value at 0.3 kOe as compared to that at zero field below 0.75 K. An increase in the critical current at 0.3 kOe, as compared to the zero field, is also found, as shown in the inset. These results suggest that the anti-proximity effect is a consequence of the strong coupling between the bulk superconducting reservoirs and the 1d ZNW's. When the coupling is weakened by the high interface resistance, the effect is weakened.

It is an interesting question to know if the effect is present in a system where the BS and nanowires are of the same material. Due to the experimental difficulties, we have not been able to make measurements on ZNW's with bulk Zn as the electrodes. However, measurements are made on Sn nanowires with diameters of 40nm (length = 6 $\mu$m) and 20nm (length = 30 $\mu$m) with bulk Sn electrodes. No evidences of anti-proximity effect are found in these samples. We have not been able to extend the measurements to sample with thinner and shorter Sn wires. A quantitative understanding of this phenomenon will require a realistic model of the superconducting order parameter in the 1-d limit and how it is altered when it is coupled to neighboring bulk superconducting reservoirs.

We acknowledge fruitful discussions with J. K. Jain, P. A. Lee, T. E. Mallouk, M. Tinkham, and X. G. Wen. This work is supported by the Center for Nanoscale Science (Penn State MRSEC) funded by NSF under grant DMR-0213623.